\documentclass[aps,prb,twocolumn,superscriptaddress,showpacs,floats,preprintnumbers,amsmath,amssymb]{revtex4}

\usepackage{graphicx}% Include figure files
\usepackage{dcolumn}% Align table columns on decimal point
\usepackage{bm}% bold math

\usepackage[pdffitwindow=false,bookmarks=true,pdfauthor={Matera et al.},colorlinks=true,citecolor=blue,linkcolor=blue]{hyperref}
\usepackage{hypernat}
\usepackage{hypcap}

\begin{document}

\preprint{(submitted to Phys. Rev. B)}

\title{Towards a first-principles chemical engineering:\\
Transport limitations and bistability in {\em in situ} CO oxidation at RuO$_2$(110)}

\author{Sebastian Matera}
\affiliation{Fritz-Haber-Institut der Max-Planck-Gesellschaft,
Faradayweg 4-6, D-14195 Berlin, Germany}

\author{Karsten Reuter}
\affiliation{Fritz-Haber-Institut der Max-Planck-Gesellschaft,
Faradayweg 4-6, D-14195 Berlin, Germany}
\affiliation{Department Chemie, Technische Universit{\"a}t M{\"u}nchen, Lichtenbergstr. 4, D-85747 Garching, Germany}

\received{June 1st, 2010}

\begin{abstract}
We present a first-principles based multiscale modeling approach to heterogeneous catalysis that integrates first-principles kinetic Monte Carlo simulations of the surface reaction chemistry into a fluid dynamical treatment of the macro-scale flow structures in the reactor. The approach is applied to a stagnation flow field in front of a single-crystal model catalyst, using the CO oxidation at RuO$_2$(110) as representative example. Our simulations show how heat and mass transfer effects can readily mask the intrinsic reactivity at gas-phase conditions typical for modern {\em in situ} experiments. For a range of gas-phase conditions we furthermore obtain multiple steady-states that arise solely from the coupling of gas-phase transport 
and surface kinetics. This additional complexity needs to be accounted for when aiming to use dedicated {\em in situ} experiments to establish an atomic-scale understanding of the function of heterogeneous catalysts at technologically relevant gas-phase conditions.
\end{abstract}

\pacs{82.65.+r,68.43.Bc,82.20.Wt,47.11.-j}

% 82.65.+r  Surface and interface chemistry; heterogeneous catalysis at surfaces
% 68.43.Bc: Ab initio calculations of adsorbate structure and reactions
% 47.11.-j 	Computational methods in fluid dynamics
% 82.20.Wt 	Computational modeling; simulation (chemical dynamics and kinetics)

\maketitle

\section{Introduction}

First-principles kinetic Monte Carlo (1p-kMC) simulations have evolved into an important tool in the modeling of heterogeneous catalytic processes \cite{reuter10}. The success of the approach relies on the accurate treatment of two central aspects for the reactive surface chemistry: A first-principles description of the involved elementary processes and an evaluation of their statistical interplay that fully accounts for the correlations, fluctuations and spatial distributions of the chemicals at the catalyst surface. Particularly if suitably combined with sensitivity analyses \cite{meskine09}, 1p-kMC simulations thus offer the prospect of an error-controlled and quantitative microkinetic modeling of the surface catalytic function. In particular for technologically relevant environments, i.e. at near-ambient reactant pressures and elevated temperatures with concomitant higher product formation rates, a third aspect comes into play that needs to be accounted for to properly describe the observable catalytic conversions. This is the flow of heat and mass in the given reactor geometry, for instance the transport of formed products away from the active surface and how efficiently the large amount of heat generated by the exothermic surface reactions can dissipate into the system. 

Corresponding macro-scale flow structures are suitably described at the continuum-level by the transient Navier-Stokes equations together with energy and species governing equations. The methodological objective of the present work is then to couple 1p-kMC into such a fluid dynamical (FD) framework, thereby augmenting the accurate treatment of the reactive surface chemistry provided by the prior technique with the capability to account for heat and mass transport effects. With a brief account of the main results already given in ref. \onlinecite{matera09}, we focus here in particular on a detailed description of this methodology. While the presented approach can readily be coupled with any computational fluid dynamics (CFD) software enabling the treatment of arbitrary reactor geometries, we develop it in the following specifically for a stagnation flow field in front of a flat-faced model catalyst. We argue that this is a suitable, though admittedly simplified reactor geometry to qualify transport effects in modern {\em in situ} studies aiming at an atomic-scale understanding of the catalytic function of single crystals in technologically relevant environments \cite{stierle07}.

The focus of such studies lies on possible differences in the surface chemistry compared to operation in ultra-high vacuum (UHV), where the function of model catalysts has been extensively studied in the past. In order to discern corresponding so-called "pressure gap" effects, it is important to assess if heat and mass transfer effects noticeably mask the true intrinsic reactivity in the {\em in situ} conditions. Particularly for the often studied CO oxidation reaction at late transition metal catalysts there are good reasons to suspect that transport limitations might not be negligible. Typically reported activities indicate a high rate of mass conversion at the surface concomitant with a large heat release. Using the established 1p-kMC model for the CO oxidation at RuO$_2$(110) \cite{reuter04,reuter06} as a representative example we illustrate with the coupled 1p-kMC--FD approach that the peculiarities of the single-crystal reactor geometry lead indeed readily to heat dissipation and mass transport limitations that severely affect the observable catalytic function. Key factors are the degree of heat conduction at the backside of the thin single-crystal, and the propensity to build-up a product boundary layer above the flat-faced surface. Obivously, such reactor-dependent effects need to be disentangled, understood and controlled when aiming to compare data obtained by different experimental setups, and when aiming to conclude on the actual surface chemistry at technologically relevant gas-phase conditions.

\section{Macro-Scale Flow Structures: Continuum Fluid Dynamics}

\begin{figure}
\centering
\includegraphics[width=7.6cm]{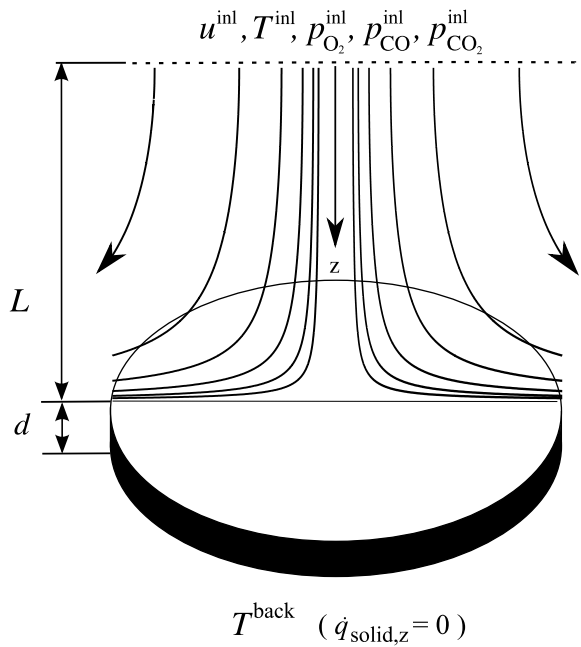}
\caption{Schematic view of the stagnation flow geometry: The gas streams from the inlet (dashed line) towards the flat-faced model catalyst of thickness $d$ and positioned at a distance $L$ away from the inlet. At the inlet the partial pressures $p^{\rm inl}_\alpha$, temperature $T^{\rm inl}$ and axial velocity $u^{\rm inl}$ are controlled; the radial velocity $v^{\rm inl}$ is zero. At the backside of the catalyst either the temperature $T^{\rm back}$ is controlled (leading here to the isothermal limit) or the heat flux $\dot{q}_{{\rm solid},z}$ is suppressed (leading to the adiabatic limit).
\label{fig1}}
\end{figure}
 
We develop our approach for a simple reactor geometry suitable to discuss heat and mass transfer effects at a flat-faced model catalyst. In the so-called stagnation flow geometry \cite{kee03} shown in Fig. \ref{fig1} the gas mixture enters through an inlet at a macroscopic distance $L$ away from the surface. At this inlet the flow is directed towards the catalyst surface, with no variation of gas composition, velocity, and temperature in the direction perpendicular to the flow. As schematically indicated in Fig. \ref{fig1} the advantage of such a geometry is that it results in an axisymmetric flow profile. Neglecting edge effects, i.e. the finite lateral extension of the model catalyst, this flow profile can effectively be described by a one-dimensional boundary-value problem. As further detailed below, this eases the analysis of the influence of the reactor setup significantly and allows to extract the relevant physics without being riddled by algebra and numerics. 

However, it is not only this symmetry imposed simplification which makes the stagnation flow setup appealing or better its realization desirable. In the spirit of the Surface Science approach to heterogeneous catalysis, the fundamental objective of {\em in situ} studies of single crystals is to obtain insight into their intrinsic reactivity in near-ambient environments, for example as function of temperature $T$ and reactant partial pressures $p_{\alpha}$. For this to be well-defined, a central prerequisite is that all, or at least a dominant fraction of the active surface sees the same gas phase. If one considers e.g. a flow geometry where the stream of reactants would approach the surface from the side -- thus in some sense an opposite 
scenario to the here discussed perpendicular stagnation flow -- then this is clearly not the case. Due to the on-going conversion of reactants into products the gas-phase composition close to the surface would gradually change across the lateral extension of the single crystal. If there are non-negligible heat transfer limitations, this goes hand in hand with a non-uniform temperature profile parallel to the surface. Under such conditions, making a defined assignment of observed turnovers to specific pressures and temperatures becomes essentially intractable. In contrast, in the stagnation flow geometry at least the entire center part of the active surface sees the same gas phase, thereby facilitating the analysis.

In the stagnation flow setup shown in Fig. \ref{fig1} the most relevant spatial coordinate is thus the direction $z$ perpendicular to the catalyst surface. In the axisymmetric problem we denote the other, radial coordinate with $r$. Using the inlet height as zero reference for $z$, the catalyst surface is then at $z = L$ and with a thickness $d$ of the single crystal its backside is located at $z = L+d$. In a continuum mechanical description the system is correspondingly characterized by two spatial regions, the flow chamber ($0<z<L$) and the sample ($L<z<L+d$), as well as three important interfaces, the inlet ($z=0$), the surface ($z=L$) and the catalyst backside ($z=L+d$). 

In the following subsections we will successively describe the modeling carried out for each of these regions and interfaces. In this manuscript this modeling aims at a description of steady-state operation, in which chemicals get converted at the active surface at a stable, time-independent rate. This rate is the turnover frequency (TOF) in units of molecules per surface area and time. We note that this time-independent formulation is a convenience, not a necessity. In fact, in particular the coupling scheme integrating the surface chemistry into the FD environment is also applicable to transient problems and we will briefly mention routes in this direction throughout the manuscript. In the same spirit it is clear that the approach is, of course, not restricted to the simple CO oxidation reaction, on which we will concentrate from now on for clarity.

\subsection{Interface I: Inlet}

At the inlet the gas flow is fully controlled. For the FD description this defines boundary values for the temperature $T(z=0) = T^{\rm inl}$, the partial pressures $p_{\alpha}(z=0) = p^{\rm inl}_{\alpha}$ (with $\alpha =$\,O$_2$, CO, CO$_2$), the total pressure $p^{\rm inl} = \sum p^{\rm inl}_{\alpha}$, and the axial velocity $u(z=0) = u^{\rm inl}$. We set the radial velocity to $v(z=0) = 0$ as is the case in stagnation flow reactors with a sieve-like showerhead as inlet \cite{kee03}, and we restrict our attention to flow situations with no circumferential motion. Instead of the partial pressures it is more common to use the total density $\rho$ and mass fractions $Y_{\alpha}$ as independent fields. For the present catalytic context the conversion is readily achieved through the ideal gas law 
\begin{equation}
p_{\alpha} \;=\; \frac{Y_{\alpha}}{m_\alpha} \rho k_{\rm B} T \quad , \label{StagIdeal}
\end{equation}
where $m_{\alpha}$ is the mass of species $\alpha$ and $k_{\rm B}$ is the Boltzmann constant. Similarly, we will see below that it is more useful to consider the scaled radial velocity $V = v/r$ in the stagnation flow equations.

\subsection{Region I: Flow chamber}

\subsubsection{Flow equations}

The continuum mechanical description of the heat and mass transport in the flow chamber is based on balance equations for total mass, species mass, momentum and internal energy. For the present purposes, the general formulation is much simplified by treating the gas phase as a Newtonian fluid with vanishing bulk viscosity, by the absence of relevant gravitational forces and by the absence of significant gas-phase chemical conversions in the context of low temperature CO oxidation (i.e. the reactions are confined to the catalyst surface). Another significant simplification arises for the laminar flows of interest here in that we can work in the low Mach number approximation: For the corresponding small flow velocities the variations of the total pressure over the whole flow chamber domain will be much smaller than its absolute magnitude. Gradients of this small variation, denoted as dynamic pressure $\hat{p}$, can then be neglected in all equations except the momentum balance.

Working in the low Mach number approximation the specific equations for the steady-state stagnation flow problem are e.g. discussed in detail by Kee, Coltrin and Glarborg\cite{kee03}. The key assumption in the derivation is that variations of partial pressure, temperature and axial velocity in radial direction are much smaller than in axial direction, at least near the symmetry axis in the center of the catalyst. A lowest order expansion in the radial dependence leads then to a set of differential equations in which all fields depend only on the axial coordinate $z$.
\begin{gather}
\intertext{Mass}
 \frac{d}{dz}(\rho u) = - 2 \rho V \label{StagMass}
\intertext{Species mass}
 \rho u \frac{d Y_\alpha}{dz} = -\frac{d j_{\alpha,z}}{dz} \label{StagSpecies}
\intertext{Axial momentum}
 \rho u \frac{d u}{dz} = -\frac{d \hat{p}}{dz} + \frac{4}{3}\frac{d}{dz}\left[ \mu \frac{du}{dz} - \mu V\right] + 2\mu \frac{dV}{dz} \label{StagAxialMom}
\intertext{Radial momentum}
 \rho u \frac{d V}{dz} + \rho V^2= -\Lambda_r + \frac{d}{dz} \left( \mu \frac{dV}{dz} \right) \label{StagRadialMom}
\intertext{Internal energy}
\rho c_p u \frac{d T}{dz} = \frac{d}{dz}\kappa \frac{d T}{dz} - \sum\limits_{\alpha} c_{p,\alpha} j_{\alpha,z} \frac{d T}{dz} \label{StagEnergy}
\end{gather}
Here, $\Lambda_r=\frac{1}{r}\partial_r \hat{p} = {\rm const.}$ is the so-called radial pressure curvature, $\mu$ the shear viscosity, $\kappa$ the thermal conductivity, $c_p$ the specific heat capacity, $c_{p,\alpha}$ the specific heat capacity of species $\alpha$, and $j_{\alpha,z}$ the axial component of the diffusive mass flux of species $\alpha$.

Together with the ideal gas law, eq. (\ref{StagIdeal}), and in the low-Mach-number-limit the condition 
\begin{equation}
\sum\limits_\alpha \frac{Y_{\alpha}}{m_{\alpha}} \rho k_{\rm B} T = {\rm const.} \equiv p^{\rm inl} \quad ,
\label{machcond}
\end{equation}
this set of equations allows to solve for all dependent fields, $\rho(z)$, $Y_\alpha(z)$, $u(z)$, $V(z)$, $j_{\alpha,z}(z)$, $T(z)$ and $\hat{p}(z)$. In fact, the axial momentum balance, eq. (\ref{StagAxialMom}), is decoupled from the other equations. All fields except the dynamic pressure $\hat{p}$ can be determined without it, and it can therefore be used afterwards to fix $\hat{p}$ from the other calculated fields. The problem gets fully determined by boundary conditions at the inlet and at the surface. Specifically, the second order equations demand independent information about $V$, $T$ and $Y_{\alpha}$ at both ends of the domain, and the first-order continuity equation requires information about $u$ on one boundary. As this information about $u$ is provided at both boundaries in the here discussed finite-gap stagnation flow, i.e. at inlet and surface, the resulting overdetermination is resolved by treating the unknown $\Lambda_r$ as an eigenvalue of the problem, i.e. its magnitude is adjusted to satisfy the additional boundary condition.

\subsubsection{Thermophysical and transport parameters}

\begin{table}
\caption{\label{tableI}
Material parameters for the gas-phase species required for the FD modeling: Characteristic diameter $\sigma$ and energy $\epsilon$, as well as vibrational frequencies $\omega$ and zero Kelvin component of the specific enthalpy $h^\circ_\alpha$ (which includes the zero-point energy contribution). The latter is referenced as usual with respect to the standard state of the atomic species, i.e. gas-phase oxygen and solid graphite for O and C, respectively.}
\begin{ruledtabular}
\begin{tabular}{ll|cccc}
                    &          & O$_2$           & CO           & CO$_2$            &\\[0.1cm] \hline 
$\sigma$            & ({\AA})  & 3.458           & 3.652        & 3.769             & [\onlinecite{cloutman00}]\\
$\epsilon/k_{\rm B}$& (K)      & 107.4           & 98.1         & 245.3             & [\onlinecite{cloutman00}]\\    
$\hbar \omega$      & (meV)    & 196             & 269          & 291, 167, 83 (2x) & [\onlinecite{webbook}]\\
$h^\circ$           & (eV)     & 0               & -1.179       & -4.074            & [\onlinecite{JANAF}] \\
%%SM Genau so. Nur mit umgekehrten Vorzeichen. Wir haben also -4.074 + 1.179= -2.895 eV (negative Reaktionenthalpie=exotherm)
%%SM PBE ueberschaetzt den Absolutwert der Reaktionsenthalpie um ca. 10% (Vergleich mit der Computational chemistry benchmark Database 
%%SM http://cccbdb.nist.gov/)
\end{tabular}
\end{ruledtabular}
\end{table}
What remains for the numerical solution are values for the transport coefficients $\mu$ and $\kappa$, and a diffusion theory relating the diffusive mass flux $j_{\alpha,z}$ to composition gradients. Further, we need expressions for the isobaric specific heat capacities $c_{p,\alpha}$ and $c_p$, as well as for the specific enthalpy $h_{\alpha}$ (see section IIC below). For the gas phase considered here the mixture specific heat is simply the mass-weighted sum of the species specific heats
\begin{equation}
c_p = \sum\limits_\alpha Y_\alpha c_{p,\alpha} \quad .
\end{equation}
For the $c_{p,\alpha}$ themselves we consider translational, rotational and vibrational degrees of freedom, treating the prior two in the classical limit and the latter in the harmonic approximation\cite{bird60}
\begin{equation}
c_{p,\alpha} = \left( \frac{3 + N^{\rm rot}_\alpha}{2}  + \sum_{i}^{N^{\rm vib}} x_{\alpha,i} \frac{e^{x_{\alpha,i}}}{\left(e^{x_{\alpha,i}} - 1 \right)^2 } \right) \frac{k_{\rm B}}{m_\alpha} \quad ,
\end{equation}
with $x_{\alpha,i} = \frac{\hbar \omega_{\alpha,i}}{k_{\rm B}T}$, and $N^{\rm vib}_{\alpha}$ and $N^{\rm rot}_{\alpha}$ the number of vibrational and rotational degrees of freedom, respectively. For the three linear molecules O$_2$, CO, and CO$_2$ $N^{\rm rot}_{\alpha} = 2$, and Table \ref{tableI} lists their vibrational frequencies $\omega_{\alpha,i}$ as taken from ref. \onlinecite{webbook}.

Computing the transport coefficients for a multi-component gas phase in complete generality is a very complex task of its own. For the accuracy level and gas-phase conditions of interest for our study effective semi-empirical molecular transport models provide fortunately a fully sufficient description. From the manifold of such existing models we adopt a strategy described by Cloutman that is also used in the reactive flow CFD program COYOTE\cite{cloutman00}. In a first step, this strategy relies on standard mixture-averaged approaches that relate $\mu$ and $\kappa$ of the multi-component gas mixture to the pure species values. In the case of the viscosity, this is the Wilke formula \cite{cloutman00,bird60}
\begin{gather}
 \mu = \sum\limits_\alpha \frac{X_\alpha \mu_\alpha}{\sum\limits_\beta X_\beta \Phi_{\alpha\beta}} \quad ,
\intertext{with}
 \Phi_{\alpha\beta} = \frac{(1+ (\frac{\mu_\alpha}{\mu_\beta})^{1/2}(\frac{m_\beta}{m_\alpha})^{1/4})^2}{8^{1/2}(1+\frac{m_\alpha}{m_\beta})^{1/2}} \quad ,
\end{gather}
where $m_\alpha$ is the mass of species $\alpha$ and $\mu_\alpha$ is the pure species viscosity. $X_\alpha$ is the molar fraction of species $\alpha$. For the thermal conductivity the analogue relation is \cite{cloutman00,bird60}
\begin{equation}
\kappa = \sum\limits_\alpha \frac{X_\alpha \kappa_\alpha}{\sum\limits_\beta X_\beta \Phi_{\alpha\beta}} \quad ,
\end{equation}
where $\kappa_\alpha$ are the pure species conductivities. Note that $\Phi_{\alpha\beta}$ is exactly the same as in Wilkes formula, i.e. it depends on the viscosities and not on the thermal conductivities. From the kinetic theory of dilute gases the expression \cite{cloutman00,bird60}
\begin{equation}
\mu_{\alpha} = \frac{5k_{\rm B}}{16 \sqrt{\pi}} \;\; \frac{m_\alpha T}{\sigma_\alpha^2 \Omega^{(2,2)*}(T^*_\alpha)} \quad ,
\end{equation}
with $T^*_\alpha = k_{\rm B}T/\epsilon_\alpha$ and
\begin{equation}
\Omega^{(2,2)*}(T^*_\alpha) = 1.147 (T^*_\alpha)^{-0.145} + (T^*_\alpha + 0.5)^2 \quad ,
\end{equation}
provides the species viscosities in terms of two empirical parameters, a characteristic diameter $\sigma_\alpha$ and a characteristic energy $\epsilon_\alpha$. For a gas with Lennard-Jones interaction between the molecules $\sigma_\alpha$ and $\epsilon_\alpha$ are the two parameters defining the interaction potential \cite{kee03}. 
For the general case referencing to a Lennard-Jones model just provides a convenient way of representing the temperature dependence of the transport coefficients and the two parameters need not to have a microscopic meaning. Values for these parameters for a wide range of species are found in data bases and we list in Table \ref{tableI} those for the species O$_2$, CO, CO$_2$ needed here\cite{cloutman00}. From the thus determined viscosities the thermal conductivities are derived via the Eucken correction\cite{cloutman00,bird60}
\begin{equation}
\kappa^\alpha = (c_{p,\alpha} + \frac{5}{4}\frac{k_{\rm B}}{m_\alpha} )\mu_\alpha \quad.
\end{equation}

In the present application the remaining diffusive mass fluxes are predominantly driven by the concentration gradients and can then be implicitly obtained from corresponding Stefan-Maxwell equations \cite{kee03,bird60} at each point in the flow field
\begin{equation}
\sum\limits_\beta \frac{k_{\rm B}T}{D_{\alpha \beta}^\text{bin}} \left[ \frac{X_\alpha}{m_\beta} j_{\beta,z} - \frac{X_\beta}{m_\alpha} j_{\alpha,z}  \right] = p^{\rm inl} \frac{dX_\alpha}{dz} \quad , \label{StefanMaxwellEqn}
\end{equation}
where the $D_{\alpha \beta}^\text{bin}$ are the binary diffusion coefficients between species $\alpha$ and $\beta$. For their determination we employ again an expression from the kinetic theory of dilute gases \cite{cloutman00,bird60}
\begin{equation}
D_{\alpha \beta}^\text{bin} = \frac{3}{16} \left(\frac{2k_{\rm B}^3}{\pi} \right)^{1/2} \;\; \frac{\left[ T^3 \left(\frac{m_\alpha + m_\beta}{m_\alpha m_\beta}\right) \right]^{1/2} }{p \sigma_{\alpha\beta}^2 \Omega^{(1,1)*}(T^*_{\alpha\beta})} \quad ,
\end{equation}
with $\sigma_{\alpha\beta} = (\sigma_\alpha + \sigma_\beta)/2$, $T^*_{\alpha\beta} = k_{\rm B}T/\sqrt{\epsilon_\alpha \epsilon_\beta}$ and
\begin{equation}
\Omega^{(1,1)*}(T^*_{\alpha\beta}) = 1.0548 (T^*_{\alpha\beta})^{-0.15504} + (T^*_{\alpha\beta} + 0.55909)^{-2.1705} \quad ,
\end{equation}
which also needs only the empirical characteristic diameter $\sigma_\alpha$ and energy $\epsilon_\alpha$ of each species, cf. Table \ref{tableI}.

\subsection{Interface II: Surface}

As already mentioned, we need to specify a number of boundary conditions at the solid surface to fully determine the set of stagnation flow equations. For the radial fluid velocity this is the standard no-slip condition, i.e. $v(z=L) = 0$. The normal component of the velocity at the surface ($u(z=L)$ in our case) is commonly termed "Stefan velocity" \cite{kee03,deutschmann08}. It is calculated by considering the mass balance at the surface\cite{mueller85}, and a finite Stefan velocity accounts for transient storage or release of species due to a changing average surface composition. For the here discussed steady-state operation this is not the case and we have $u(z=L) = 0$. 

As the CO oxidation reactions at the surface are the only processes that consume reactants and yield products in stationary operation, the boundary condition for the diffusive mass flux is
\begin{equation}
j_{\alpha,z}(z=L) \;=\; -\tau_\alpha^{\rm surf} \;=\; - m_\alpha \nu_\alpha {\rm TOF} \quad ,
\label{chemsource}
\end{equation}
where the chemical source term $\tau_\alpha^{\rm surf}$ for each species is simply given by the overall rate of reaction events (per area and time) times the stoichiometry coefficient $\nu_\alpha$ and mass $m_\alpha$ of the species in the reaction. For the simple CO oxidation reaction, $\nu_{\rm CO} = -1$, $\nu_{\rm O_2} = -1/2$, and $\nu_{\rm CO_2} = 1$. 

The heat release connected to the exothermic conversions must also enter the heat balance at the surface. With the previously introduced boundary conditions the surface energy balance reduces to the mere requirement that the normal heat flux is continuous at the surface
\begin{equation}
- \kappa \frac{dT}{dz}(z=L) + \sum\limits_{\alpha} h_{\alpha} j_{\alpha,z}(z=L) \;=\; \dot{q}_{{\rm solid},z}(z=L) \quad .
\label{surfbalance}
\end{equation}
Here, the two terms on the left hand side account for the heat transported away by the gas phase and the heat released by the surface chemical reactions, respectively, which must balance the heat flux into the solid $\dot{q}_{{\rm solid},z}(z=L)$. The temperature-dependent contribution to the specific enthalpies $h_\alpha$ is determined completely equivalently to the specific heat capacities, i.e. by considering translational, rotational and vibrational degrees of freedom as described above. The additionally required zero Kelvin component $h^\circ_\alpha$ including the vibrational zero point energies can be drawn from thermochemical tables \cite{JANAF}, and for completeness they are included in Table \ref{tableI} for the species O$_2$, CO, and CO$_2$.

At first glance this use of experimental thermochemical data might seem inconsistent with the use of first-principles energetics in the 1p-kMC simulations. Our choice is motivated by the consideration that all other transport parameters are equally derived from experiment. The empirical transport models provide a fully sufficient and controlled description of the macro-scale flow structures at the accuracy level of interest for this study. As argued in more detail below such a description is reached for the reactive surface chemistry through the use of first-principles based 1p-kMC modeling. We are thus faced with the problem of matching these two descriptions. In our approach this match occurs uniquely through the TOFs. Whatever energetics is required for their determination comes exclusively and consistently from first-principles, here density-functional theory with a semi-local exchange-correlation functional. Vice versa, the entire transport description is exclusively and consistently based on experimental numbers, such that the balance leading to its effective accuracy is not disturbed by occasional parameters as e.g. the $h^\circ_\alpha$ coming from approximate first-principles theory. While we feel that such considerations are necessary for the envisioned error-controlled multiscale modeling, we note that in practice the semi-local functional employed in the 1p-kMC overestimates the zero Kelvin enthalpy change for the CO oxidation reaction in Table \ref{tableI} by only about 10\%, such that none of the conclusions reported below would be touched when using this energetics instead of the experimental one.

Since we are considering heat transport in both the solid and the gas phase, we finally need two conditions describing the change of the temperature field when crossing the surface. Equation (\ref{surfbalance}) can only serve as boundary condition for the gas-phase heat transport. Another one is needed for the heat transport in the solid. Here, we want to assume that the temperature is continuous across the gas-solid interface. Within our interest in modeling the reactivity at near-ambient conditions, this should be rather well ensured by the frequent gas-surface collisions.

\subsection{Region II and Interface III: Sample and sample backside}

Neglecting possible heat-induced deformations or phase transformations we only account for heat conduction through the single crystal. 
For the stagnation flow problem there is no radial variation of the gas-phase temperature. 
We maintain this description within the sample and correspondingly also model the heat transport as a one-dimensional problem described by Fourier's law
\begin{equation}
\dot{q}_{{\rm solid},z} = - \kappa_{\rm solid} \frac{dT}{dz} \quad ,
\label{heatrans}
\end{equation}
where $\kappa_{\rm solid}$ is the heat conductivity of the sample, which for simplicity we assume to be temperature-independent. Solution of this equation requires fixing a boundary value at the backside of the single crystal $\dot{q}_{{\rm solid},z}(z=L+d)$, which thus describes the degree of heat dissipation that is possible e.g. through radiative loss or the contact of the single crystal with the sample holder. Specifying this for real experimental setups is a demanding task and we suspect that in present {\em in situ} experiments this value will vary largely. Addressing these specificities in a quantitative way is clearly outside the capabilities of the present idealized reactor model. With real experimental setups lying anywhere in between we therefore analyze the relevance of this factor for thin single crystals by considering two opposite extremes: First a fixed temperature at the sample backside to mimic a highly efficient heat coupling of the crystal to the system, and second a zero heat flux at the sample backside to represent a well insulated sample. For either boundary condition eq. (\ref{heatrans}) can be solved analytically and provides in turn the missing boundary condition for the surface heat balance, eq. (\ref{surfbalance}).
\begin{gather}
\intertext{Isolated sample (adiabatic limit)}
\dot{q}_{{\rm solid},z}(z=L) \;=\; 0 \quad .
\intertext{Connected sample (isothermal limit)}
\dot{q}_{{\rm solid},z}(z=L) \;=\; \frac{\kappa_{\rm solid}}{d} \left( T(z=L)-T^{\rm back} \right) \quad ,
\end{gather}
where $T^{\rm back}$ is the temperature at the back of the crystal. While in principle this temperature could have any value (e.g. through controlled heating) we will assume here that it is identical to the one of the outside system and thus to the temperature of the incoming gas at the inlet, i.e. $T^{\rm back} \equiv T^{\rm inl}$. The remaining material parameter to specify in the resulting boundary condition is the bulk heat conductivity. An experimental value for bulk RuO$_2$ is $\kappa_{\rm solid(RuO_2)}$= 0.50 W cm$^{-1}$ K$^{-1}$.\cite{ferizovic_determination_2009} However, almost all {\em in situ} work on RuO$_2$(110) has in fact been performed on ultra-thin films grown on Ru crystals, which would suggest that the value $\kappa_{\rm solid(Ru)}$= 1.17 W cm$^{-1}$ K$^{-1}$ is more appropriate \cite{Lide1995}. Fortunately, for the results reported below it makes no difference which value for this quantity is used. At the rather high thermal conductivity of either metallic Ru or the metallic oxide RuO$_2$ this dissipation channel is so dominant, that -- as soon as it is enabled in the surface heat balance, eq. (\ref{surfbalance}) -- it simply ensures that the surface temperature remains at the nominal value $T^{\rm surf}(z=L) = T^{\rm inl}$. Regardless of the specific value of $\kappa^{\rm solid}$ the second boundary condition is therefore simply equivalent to modeling an isothermal reactor limit, while the first boundary condition would correspond to the adiabatic limit.

\subsection{Numerical solution}

The stagnation flow equations, eqs. (\ref{StagIdeal}) and (\ref{StagMass}-\ref{machcond}), can be transformed into a semi-explicit system of differential-algebraic equations (DAE)
\begin{eqnarray}
\nonumber
\frac{d}{dz} y_i &=& f(y_j,C,z) \quad , \\
               0 &=& g(y_j,C,z) \quad ,
\end{eqnarray}
where $y_j$ are the so-called differential components and $C$ the algebraic component. In our case the differential components are density $\rho$, mass fractions $Y^\alpha$, temperature $T$, velocity components $u$ and $V$, diffusive mass fluxes $j_{\alpha,z}$, intrinsic heat flux $-\kappa \frac{dT }{dz}$, pressure curvature $\Lambda_r$, and $\mu \frac{dV}{dz}$. The algebraic component is the gradient of the density, which is determined by the requirement that the total pressure is constant between inlet and surface.

We solve the DAE boundary value problem numerically using the COLDAE package \cite{ascher94}. This package uses piecewise orthogonal collocation at Gaussian points and has an adaptive mesh strategy allowing for an error-controlled solution. We use the default option controlling the number of intermediate points and an initial equidistant mesh with 10 spacings. In all simulations we employ a tolerance of $10^{-4}$ for each differential component. In order to improve the stability and have full error control, all variables, dependent or independent, are presented in appropriate units. The employed units are the inlet-surface distance $L$ for length, and for the velocity, density, and mass fractions their values at the inlet. The employed temperature scale is Kelvins. We use the representation $(T-T^{\rm inl})$ for the temperature, so that this renormalized temperature is always zero at the inlet. The mass fluxes, heat flux, and density gradient are expressed in multiples of $\frac{D^\text{eff,inl} Y^\text{inl}_\alpha}{L}$, $\frac{\kappa^\text{inl}}{L}$, and $\frac{\rho^\text{inl}}{L}$, respectively, where $D^\text{eff,inl}$ is the mixture averaged diffusion coefficient\cite{kee03}. Finally, the radial pressure curvature $\Lambda_r$ and $\mu \frac{dV}{dz}$ are scaled with $100 \mu^\text{inl} \frac{u^\text{inl}}{L^3}$ and $\mu^{\rm inl}\frac{u^\text{inl}}{L^2}$, respectively.

The software uses a (damped) Newton strategy to find a solution starting from an initial guess. The central features of the initial guess for the first simulation were constant $\rho$, $T$ and $Y_\alpha$, as well as as a third order polynomial for $u$ that fulfills the boundary conditions. All remaining unknowns were approximated according to these assumptions. For subsequent simulations we used the results of previous simulations where appropriate. In those cases the adapted mesh from the previous simulation was coarsened to contain only half as many grid points as initially.

\section{Surface reaction chemistry: First-principles kinetic Monte Carlo simulations}

The actual surface catalytic activity enters the FD simulations through the TOFs in the boundary value eq. (\ref{chemsource}) for the partial mass fluxes at the surface. The corresponding calculation of the rate of product formation per surface area and time from information about the elementary processes in the catalytic cycle is the realm of microkinetic theories. The most prominent such approach relies on phenomenological rate equations which only consider the mean-field averaged concentrations (coverages) of the reaction intermediates at the surface\cite{dumesic_microkinetics_1993}. This level of modeling of the surface chemistry is the prevalent standard in reactor engineering\cite{kee03,deutschmann08}. There, the kinetic quantities entering the rate expressions are in fact often treated as adjustable parameters. In a top-down fashion the idea is thus to use macroscopic reactor data to derive effective insight into the on-going surface catalytic activity. In more bottom-up oriented work the kinetic quantities are alternatively drawn from independent detailed experiments, or in modern hybrid approaches increasingly from first-principles calculations\cite{gokhale_molecular-level_2004,maestri_c-1_2009}. The idea of such integrated approaches is correspondingly to model how both the intrinsic surface chemistry and transport effects contribute together to the macroscopically observable activity in a given reactor setup.

The latter is also the central objective of the present study. However, for the here aspired quantitative modeling present-day hybrid approaches are not sufficient. Use of scattered experimental and first-principles kinetic data from different sources and in the latter case potentially from different levels of approximate theory incurs a rather uncontrollable error. Even if the kinetic parameters of all involved elementary processes were reliable, there is still the error from the approximate mean-field treatment underlying the rate equation approach. In fact, for the here studied CO oxidation reaction at RuO$_2$(110) this error has recently been shown to be qualitative with deviations of the mean-field TOFs spanning up to several orders of magnitude\cite{temel07}. Aiming at an error-controlled multiscale modeling of predictive quality we therefore employ for the description of the surface kinetics 1p-kMC as most accurate approach with explicit account of the correlations, fluctuations and detailed spatial distributions of the chemicals at the surface.

\subsection{1p-kMC model of CO oxidation at RuO$_2$(110)}

The molecular-level basis of 1p-kMC is a microscopically correct first-principles description of the elementary processes involved in the catalytic cycle. In the established model of CO oxidation at RuO$_2$(110) \cite{reuter04,reuter06} this is specifically the set of 26 elementary processes defined by all non-correlated site and element specific adsorption, desorption, diffusion and reaction events that can occur on a lattice spanned by two different active sites offered by the surface, so-called bridge (br) and coordinatively unsaturated (cus) sites. For all these processes density-functional theory in conjunction with harmonic transition-state theory is used to compute the kinetic parameters. The resulting 26 first-principles rate constants form then the essential input for the actual 1p-kMC simulations which evaluate the statistical interplay among the surface chemical processes by following the long-term time evolution of the open catalytic system through numerical solution of a Markovian master equation\cite{reuter10}. 

Using exactly the computational setup as detailed before\cite{reuter04,reuter06,temel07}, these simulations are carried out for the present purposes for a given local temperature $T^{\rm surf}$ and reactant partial pressures $p^{\rm surf}_{{\rm O}_2}$ and $p^{\rm surf}_{\rm CO}$ directly at the surface, which in particular fixes the impingement and therewith the rate constants of the adsorption processes. Under such conditions, the system eventually reaches a {\em unique} steady-state, in which the detailed surface composition and occurrence of the individual elementary processes still exhibit the correct microscopic fluctuations, yet when averaged over simulation cells exceeding the characteristic correlation lengths at the surface have well-defined and constant values. These values thus comprise the average rate of reactant to product conversion under the given gas-phase impingement and local temperature, i.e. exactly the TOFs that enter the partial mass flux boundary condition of eq. (\ref{chemsource}). While it is only this averaged quantity that matters for the macroscopically described flow field, it is important to note that it is still properly derived from microscopic simulations that fully account for the site heterogeneity and distributions at the surface. This is thus distinctly different to the mentioned mean-field based phenomenological descriptions that are commonly integrated in the CFD modeling of macro-scale flow structures.

\subsection{Integration of 1p-kMC into the fluid dynamical environment}

The 1p-kMC and FD simulations are intricately coupled. On the one hand, the TOFs required in eq. (\ref{chemsource}) to close the stagnation flow equations are provided by the 1p-kMC simulations. On the other hand, fixing the surface impingement in the 1p-kMC simulations requires the local temperature and gas-phase partial pressures directly at the surface, which are determined by the heat and mass transport modeled at the continuum level. A straightforward approach to this interdependence is a simultaneous solution until self-consistency between flow and 1p-kMC boundary condition is achieved\cite{vlachos97}. For the here discussed stagnation flow equations this approach is in principle feasible\cite{kissel98}, albeit potentially numerically unstable\cite{chatterjee04}. However, for more complex reactor geometries it would quickly become intractable, as usually several independent 1p-kMC simulations would be required for every spatially resolved cell at the surface. Precisely due to the necessity to continuously rerun 1p-kMC simulations the approach would also be very inefficient for the here intended simulation of catalytic activity at a large variety of flow conditions.

We instead achieve a computationally much more efficient formulation by decoupling the interdependence through an instantaneous steady-state approximation \cite{deutschmann08}. The kMC simulations are first carried out to determine the steady-state TOFs for a wide range of surface impingement and local temperature conditions. The resulting grid data is then interpolated to a continuous representation, which in turn provides the entire necessary boundary condition for the stagnation flow problem. For the steady-state operation targeted in this manuscript this divide-and-conquer type approach is exact and may easily be applied to more complex reactor geometries. It could even be extended to transient phenomena under the assumption that on the time scale of relevant flux variations the surface chemistry adapts quasi instantaneously to the new steady-state, hence the name.

For the CO oxidation at RuO$_2$(110) we thus first computed 1p-kMC steady-state TOFs for the entire relevant range of temperatures and gas-phase composition. With a negligible CO$_2$ readsorption probability these TOFs are independent of the CO$_2$ partial pressure. Specifically we then used a dense grid with 25\,K spacing for the temperature range 400\,K $< T^{\rm surf} <$\,850\,K and with logscale spacing to cover the range $10^{-6}$\,atm $< p^{\rm surf}_{\rm O_2} < 10^2$\,atm with 30 and the range $10^{-5}$\,atm $< p^{\rm surf}_{\rm CO} < 10^2$\,atm with 42 spacings. Through modified quadratic Shepard interpolation\cite{renka88,renka88_2} this is converted into a reliable continuous representation TOF($T^{\rm surf}, p^{\rm surf}_{\rm O_2}, p^{\rm surf}_{\rm CO}$) that is finally presented as boundary condition to the stagnation flow solver.

\section{Results}

The intrinsic activity resulting from the 1p-kMC model for the CO oxidation at RuO$_2$(110) has been analyzed for a wide range of temperature and pressure conditions before \cite{reuter04,reuter06}. Not surprisingly, high catalytic activity is only observed for a rather narrow range of gas-phase conditions, which stabilize O and CO simultaneously at the surface in appreciable amounts. For more O-rich feed the surface is poisoned by oxygen, for more CO-rich feed the surface is poisoned by CO, and little CO$_2$ is formed in either case. For gas-phase conditions in the UHV regime, which allow direct measurements of the intrinsic activity, the model reproduces existing experimental TOF data quantitatively \cite{reuter04,reuter06}. It must be stressed though that for the description of the CO-poisoned regime the model has clear limitations. In corresponding CO-rich feeds the oxide surface would in reality eventually be reduced. By construction, this and the catalytic activity connected to such a phase transformation can not be grasped by the present 1p-kMC model assuming an intact underlying RuO$_2$(110) lattice. 

Notwithstanding, this restriction concerns the modeling of the surface chemistry. The focus of the present work is instead to quantitatively integrate a given microkinetic description of this surface chemistry into a FD framework to assess heat and mass transport effects in a reactor geometry representative for {\em in situ} studies of model catalysts. For this endeavor the existing account of the surface chemistry is  -- despite its noted limitation -- very suitable, in particular as it exhibits a number of features that we consider rather generic for a high-TOF reaction like the CO oxidation: (i) The intrinsic catalytic activity is narrowly peaked in a small range of gas-phase conditions. (ii) This activity is not sufficiently described by standard rate equation formulations. For predictive quality the first-principles based microkinetic modeling must therefore be based on an approach like 1p-kMC that explicitly accounts for the detailed spatial distribution of the chemicals at the surface. In turn, it is the latter type of approach to which the FD environment must be coupled, e.g. through the instantaneous steady-state approximation employed here. (iii) The peak activity at optimum partial pressures increases rapidly in the temperature range of interest. Towards the upper end at around 500-600\,K and together with the high exothermicity of the CO oxidation reaction, this leads to a degree of mass conversion and heat release at the surface that is prone to transport limitations in the reactor. 

As we will see the amount of heat dissipation possible at the back of the thin single-crystal is a crucial factor for such limitations that can mask the true intrinsic activity in {\em in situ} model catalyst studies. Not aiming (nor being able to) give a detailed account for one specific experimental setup we will analyze this in the following in more generic terms for the two already described opposite limits. In the adiabatic limit there is no heat flux at all through the sample backside, mimicking to some degree the situation that could e.g. arise from an insulating sample holder. In contrast, in the isothermal limit we assume the sample to be sufficiently well connected to the outside system that a constant nominal temperature is maintained throughout. Here, this is chosen to be the same temperature as that of the gases at the inlet. 

Considering the peaked structure of the intrinsic catalytic activity in $(T,p^{\rm surf}_{\rm O_2},p^{\rm surf}_{\rm CO})$-space we can conveniently study heat and mass transfer effects in these two limits focusing on prototypical sets of gas-phase conditions: For defined temperature, essentially zero CO$_2$ concentration ($p^{\rm inl}_{\rm CO_2} \equiv 10^{-5}$\,atm throughout) and near-ambient oxygen partial pressure at the inlet these sets comprise a range of inlet CO partial pressures. They cover the O-poisoned regime at the lowest $p_{\rm CO}^{\rm inl}$, the CO-poisoned regime at the highest $p_{\rm CO}^{\rm inl}$, and span over the conditions of highest intrinsic activity.

\subsection{Adiabatic limit}

\subsubsection{Surface heating}

\begin{figure}
\includegraphics[width=\linewidth]{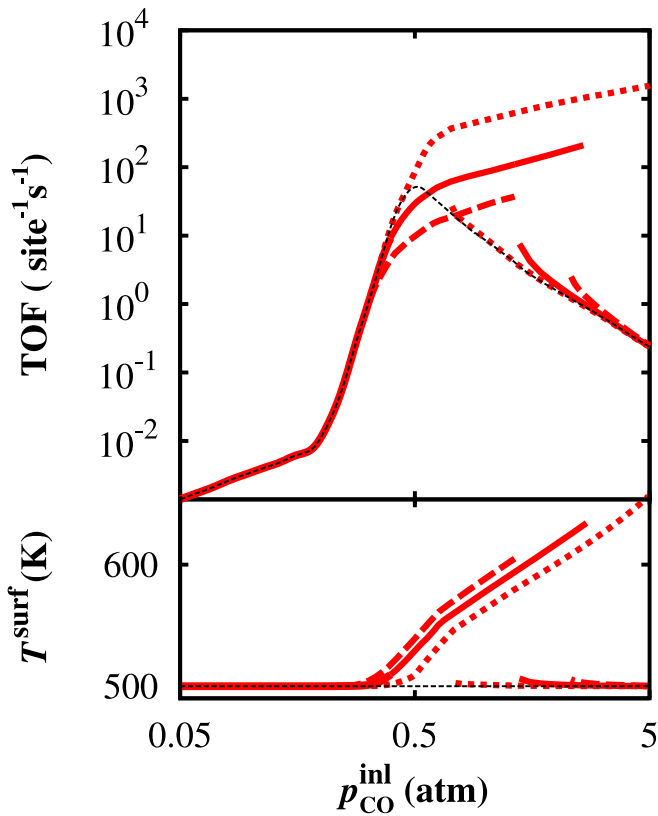}
\caption{(Color online) Comparison of intrinsic steady-state TOFs as resulting from 1p-kMC (black thin dashed line) with observable TOFs when accounting for transport effects in the stagnation flow reactor in the adiabatic limit. For $T^{\rm inl} = 500$\,K and $p^\text{inl}_\text{O$_2$} = 0.3$\,atm the shown range of inlet CO partial pressures spans from the O-poisoned to the CO-poisoned regime with the intrinsic "most active" state reached at intermediate $p^{\rm inl}_{\rm CO}$ corresponding roughly to a stoichiometric feed. The suppressed heat flux at the sample backside allows the system to sustain a high-activity operation mode for more CO-rich conditions than this nominal most active state. In this high-activity branch of the TOF-profile the surface temperature, $T^{\rm surf}$ shown in the lower panel, is significantly increased. The activity and extension of the branch depends on the reactor details. Shown here are results for constant inlet velocity $u^{\rm inl} = 1$\,cm/s and varying inlet-surface distance: $L = 1$\,mm (dotted red line), $L = 1$\,cm (solid red line) and $L = 10$\,cm (dashed red line).}
\label{fig2}
\end{figure}

The heat flux through the back of the sample is completely suppressed in the adiabatic limit. The only dissipation channel left for the heat released by the exothermic surface reactions is then into the surrounding gas phase. Compared to the heat conduction through a metallic sample this channel is rather inefficient. One can therefore suspect that it may not be efficient enough to maintain the nominal surface temperature, once the TOF and therewith the generated heat rate exceeds a certain critical value. Instead, the surface will heat up and give rise to gas-phase temperature gradients from inlet to active surface. This is indeed what we find in the coupled 1p-kMC--FD simulations. When using representative parameters for the inlet distance $L = 1$\,cm and axial inlet flow velocity $u^{\rm inl} = 1$\,cm/s significant deviations from the nominal surface temperature set in for near-ambient environments at TOFs exceeding $\sim 10$\,site$^{-1}$ s$^{-1}$. Such peak TOFs are reached for optimum partial pressure conditions at temperatures above about 500\,K. 

The effect of the ensuing temperature increase at the active surface on the observable steady-state conversion rates is quite dramatic already at this threshold temperature and is illustrated in Fig. \ref{fig2}. While the intrinsic activity is peaked close to stoichiometric feeds, the heat transport limitations lead to the stabilization of a high-activity operation mode that extends to significantly more CO-rich conditions. In this branch of the observable TOF-profile shown in Fig. \ref{fig2} the surface temperature is up to 150\,K higher than the nominal temperature at the inlet. Quantitatively this and the concomitant TOFs in the newly established high-activity mode depend on the details of the reactor setup. This is exemplified in Fig. \ref{fig2} with the TOF-profiles that result when increasing or decreasing the inlet distance by one order of magnitude. For a larger $L = 10$\,cm the extension of the high-activity branch is reduced at overall lower conversion rates, while for a smaller $L = 1$\,mm the branch extends to much higher CO partial pressures and the observable conversion rate exceeds the peak intrinsic activity by more than one order of magnitude. Similar, but quantitatively smaller variations are obtained when changing the inlet velocity by one order of magnitude up or down (not shown). For a lower $u^{\rm inl} = 1$\,mm/s the high-activity branch exhibits slightly smaller TOFs and extends up to slightly smaller $p^{\rm inl}_{\rm CO}$ than for the $u^{\rm inl} =1$\,cm/s displayed in Fig. \ref{fig2}. An increase to $u^{\rm inl} =10$\,cm/s, on the other hand, increases the extension of the branch to higher CO partial pressures at also higher TOFs than those shown in Fig. \ref{fig2}. 

Regardless of these quantitative variations, the net effect of the surface heating resulting from the transport limitations is in all cases a substantially changed observable TOF-profile compared to the true underlying intrinsic reactivity. The absolute TOFs in the relevant high-activity regime are significantly different and the inlet gas-phase conditions for which highest conversions are obtained are shifted to much more CO-rich feeds. Obviously, if these observable TOFs were mistaken for the intrinsic reactivity, wrong mechanistic conclusions about the on-going surface chemistry in such {\em in situ} environments would be derived. Furthermore, the observable TOF-profile in Fig. \ref{fig2} exhibits another feature that is completely absent in the intrinsic reactivity: For a range of CO partial pressures we obtain two stationary solutions: The high-activity branch and in addition a low-activity branch that coincides with the intrinsic activity. As the underlying 1p-kMC model of CO oxidation at RuO$_2$(110) has no multiple steady-states\cite{temel07}, this bistability arises solely from the coupling of macroscopic transport and surface chemistry.

\subsubsection{Formation of a boundary layer}

\begin{figure}
\includegraphics[width=\linewidth]{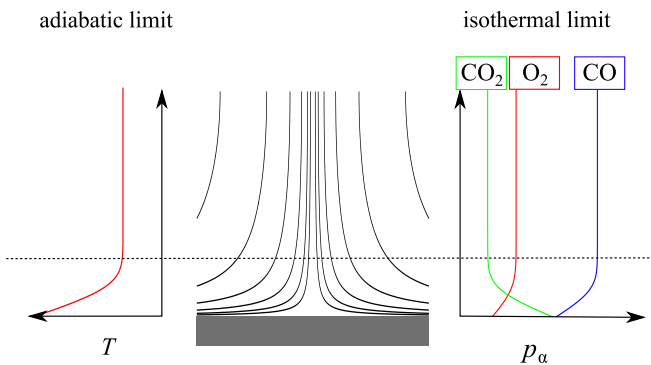}
\caption{(Color online) Illustration of the temperature and pressure profiles in steady-state stagnation flow. Diffusion and heat conduction only take place in a boundary layer above the surface; outside transport is purely convective. In the here discussed adiabatic limit sufficiently large TOFs lead to an increase of temperature in this boundary layer, while in the isothermal limit they lead to a change of partial pressures. In the general case both effects can be intricately intermingled. The boundary layer expands when the inlet velocity is decreased, ultimately filling the whole gap between inlet and surface in the limit $u^\text{inl} \rightarrow 0$. The variation with the inlet-surface distance $L$ is opposite, i.e. the boundary layer shrinks with decreasing $L$.}
\label{fig3}
\end{figure}
 
An analysis of the observed range of transport effects (high-activity branch with concomitant bistability and variations with reactor setup) starts from the anticipated formation of a finite boundary layer above the flat-faced model catalyst. As schematically drawn in Fig. \ref{fig3} non-vanishing gradients of temperature and partial pressures are in general restricted to this boundary layer. This is due to the convective nature of the transport. The gas streams towards the surface and the chemical reactions at the surface have no influence on the fields far away. Heat and species move with the flow, which dominates over any non-convective transport. Only when the axial velocity approaches zero near the surface, diffusion and heat conduction kick in, as they are now of similar size as the convective transport. For the here discussed adiabatic limit the continuous catalytic conversions at the surface lead to a continuous heat release into the gas phase. For sufficiently large TOFs above the critical value this heat rate is too high to be efficiently transported away by the flow. The result is a steady-state temperature profile above the surface as sketched in Fig. \ref{fig3}, where the temperature rises within the boundary layer from $T^{\rm inl}$ to the values for $T^{\rm surf}$ shown in Fig. \ref{fig2}. On the other hand, surface mass conversions of the order of magnitude as those of Fig. \ref{fig2} are still too low to significantly affect the gas composition in the boundary layer. The formed products are transported away sufficiently quickly so that the partial pressures remain essentially unchanged, i.e. we find $p^{\rm surf}_\alpha \approx p^{\rm inl}_\alpha$ for all conditions discussed in Fig. \ref{fig2}.

\begin{figure}
\includegraphics[width=\linewidth]{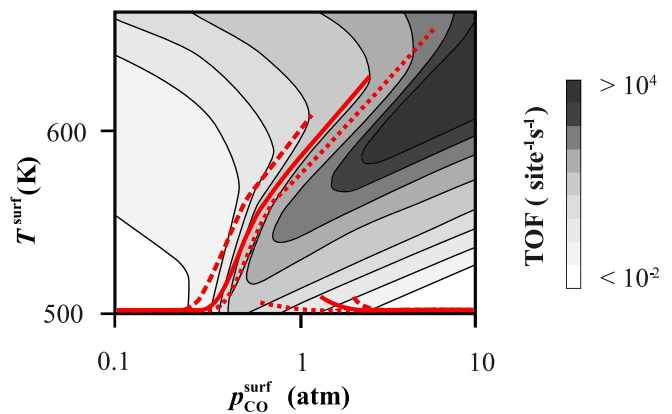}
\caption{(Color online) Intrinsic TOF contour plot as computed with 1p-kMC for the same constant $p_{\rm O_2}^{\rm inl} \approx p_{\rm O_2}^{\rm surf} = 0.3$\,atm as in Fig. \ref{fig2}. Marked by red lines are the TOF and surface temperature that result in the steady-state stagnation flow with $u^{\rm inl} = 1$\,cm/s and varying inlet distances, i.e. the conditions behind the different red lines in Fig. \ref{fig2}. The line shapes follow the ones of Fig. \ref{fig2}, i.e. $L = 1$\,mm (dotted red line), $L = 1$\,cm (solid red line) and $L = 10$\,cm (dashed red line).}
\label{fig4}
\end{figure}
 
The latter feature allows us to suitably discuss the observed transport effects in terms of the intrinsic reactivity summarized in Fig. \ref{fig4}. Shown is the TOF profile as obtained with 1p-kMC for the same $p_{\rm O_2}^{\rm inl} \approx p_{\rm O_2}^{\rm surf} = 0.3$\,atm as in Fig. \ref{fig2} and as a function of surface temperature and CO partial pressure. The region of highest-activity is narrowly peaked and the peak TOFs at this rim increase steadily with temperature. A crucial feature for the following is that the rim does not go along constant partial pressure ratio for higher temperatures (i.e. vertically up in Fig. \ref{fig4}), but shifts continuously towards more CO-rich mixtures (i.e. diagonally up in Fig. \ref{fig4}).
%%SM Wenn ich muCO/muO2 als Achse benutze, sieht fig4 fast genauso aus wie mit pCO. Nur bei muCO als Achse ist der Rim senkrecht.
This is because the most active state is characterized by the coexistence of O and CO at the surface, which follows more a constant ratio of chemical potentials than partial pressures. Imagine now that the system starts at the nominal inlet and surface temperature of 500\,K, corresponding to the bottom horizontal line in Fig. \ref{fig4}. As long as the intrinsic TOF does not exceed the critical value, the system is able to maintain this temperature and the observable TOF is identical to the intrinsic one, cf. Fig. \ref{fig2}. For the range 0.4\,atm $< p^{\rm surf}_{\rm CO} < 0.85$\,atm the activity is, however, above the critical value. The system cannot transport the generated heat away sufficiently quickly, and surface and boundary layer start to heat up. As directly apparent by focusing on e.g. $p^{\rm surf}_{\rm CO} = 0.5$\,atm in this critical pressure range the intrinsic TOFs increase with increasing surface temperature (vertically up in Fig. \ref{fig4}). This gives rise to a runaway process. Higher TOFs generate more heat, which increases the surface temperature and therewith leads to even higher TOFs. This will only stop once the system is over the highest activity rim in Fig. \ref{fig4}. Further increase in surface temperature (i.e. moving even more vertically up along the line of constant $p_{\rm CO}^{\rm surf}$ in Fig. \ref{fig4}) leads then to a decrease in intrinsic reactivity. This allows the system to find a new steady-state, viz. the high-activity branch corresponding to the increased surface temperature in Fig. \ref{fig2}. 

Where on this downward TOF slope the system precisely settles down depends on how efficiently the generated heat can be transported away, i.e. which surface temperature results for the heat rate connected with a given intrinsic TOF. This is crucially controlled by the thickness of the boundary layer and corresponding thickness variations rationalize the entire observed dependence on the reactor setup. A smaller inlet distance compresses the boundary layer and therewith enables a better heat convection. Correspondingly, at smaller $L$ the system ends up closer to the most active rim in Fig. \ref{fig4} and the high-activity operation mode exhibits higher observable TOFs as seen in Fig. \ref{fig2}. A higher inlet velocity has the same effect on the boundary layer, and therewith also leads to higher observable TOFs as described above. 

The intrinsic TOF profile shown in Fig. \ref{fig4} also helps to understand why the high-activity branches eventually break down and how their extension varies with the reactor setup. Apart from the highest-activity rim dominated by the oxidation reactions between CO and O adsorbed at the most active cus sites one can discern at the upper edge of Fig. \ref{fig4} a weak new rise of the intrinsic TOFs. At these highest temperatures shown this is due to the "entropic" widening of the kinetic "phase" transition region where the appropriate chemical potential ratio in the gas-phase stabilizes the coexistence of both reactants at the surface \cite{reuter06,reuter03}. When over the rim the hitherto described decrease of the intrinsic TOF with increasing temperature will therefore eventually change into a new increase. If the system reaches this change of slope, a new runaway cycle of increasing temperature and TOF will start and no stationary operation mode can be stabilized (at least not in the temperature range of interest for the present study). Correspondingly, in Fig. \ref{fig4} the high-activity branches always break down at positions of such a gradient change in the intrinsic TOF profile. As apparent from Fig. \ref{fig4} the further away from the rim the high-activity branch is situated, the earlier it hits this slope change, i.e. its extension reaches only up to smaller $p^{\rm inl}_{\rm CO}$. In the present adiabatic limit, modifications of the reactor setup that compress the boundary layer (either by higher axial inlet velocity or smaller inlet distance) lead therefore to higher absolute observable TOFs in a high-activity branch that extends up to higher CO partial pressures.

Finally, some remarks about the observed bistability are at place. The structure of the TOF profile in Fig. \ref{fig4} rationalizes why for some pressure conditions two steady-state solutions are obtained. One, in which the system exhibits a low activity that coincides with the intrinsic one, and one, in which significant surface heating has brought the system above the highest-activity rim. While intuitive, the rationalization in terms of thermal runaway is at present clearly an interpretation. A verification would require the extension of the present steady-state approach to transient operation. Only corresponding time-resolved simulations will then give access to the wealth of phenomena that are now only suggested by the observed bistability. Notably this is the possibility of oscillations between the two modes. In contrast to e.g. purely surface reaction--surface diffusion driven oscillations on single-crystals in UHV \cite{imbihl09} the mechanistic details behind corresponding reactor--reaction oscillations in the ambient pressure regime are only poorly understood \cite{schueth93}. Obviously, extending the present model in this direction offers the prospect of a detailed analysis, on which we will concentrate in future work.

\subsection{Isothermal limit}

\begin{figure}
\includegraphics[width=\linewidth]{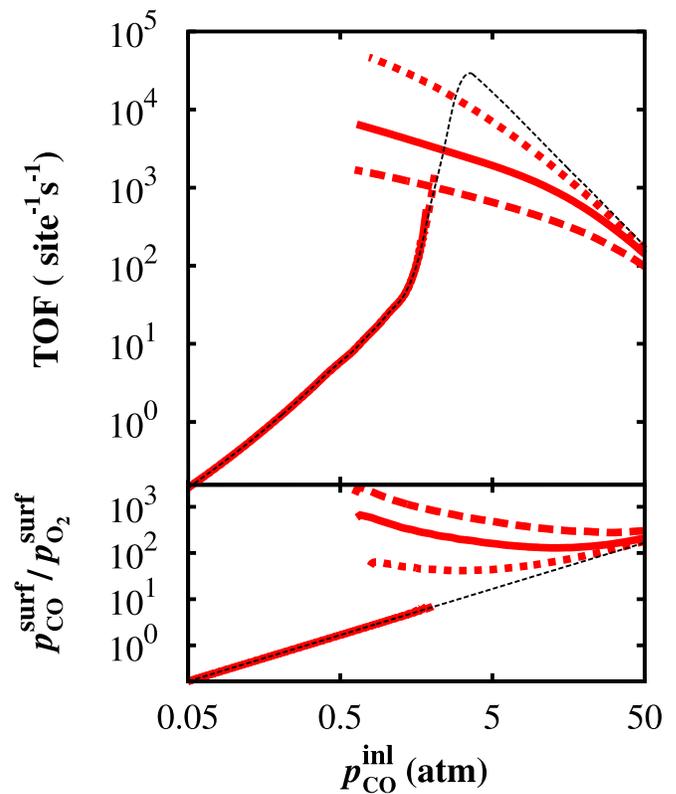}
\caption{(Color online) Comparison of intrinsic steady-state TOFs as resulting from 1p-kMC (black thin dashed line) with observable TOFs when accounting for transport effects in the stagnation flow reactor in the isothermal limit. As in Fig. \ref{fig2} the data is for $p^\text{inl}_\text{O$_2$} = 0.3$\,atm, but now for a higher temperature $T^{\rm inl} = 600$\,K. The higher intrinsic peak TOFs at this higher temperature are significantly masked by mass transfer limitations. The variation with the reactor setup is illustrated for constant inlet velocity $u^{\rm inl} = 1$\,cm/s by varying inlet-surface distances, $L = 1$\,mm (dotted red line), $L = 1$\,cm (solid red line) and $L = 10$\,cm (dashed red line). Using the same line styles the lower panel shows how the partial pressure ratio at the surface deviates from the nominal one at the inlet.}
\label{fig5}
\end{figure}

In the opposite isothermal limit the high thermal conductivity of the metallic sample allows for such an efficient removal of the generated heat that even at higher temperatures around 600\,K, where the intrinsic peak TOFs at optimum partial pressures exceed $10^4$\,site$^{-1}$s$^{-1}$, no significant surface heating results. The temperature remains at the nominal inlet value throughout the entire system. As shown in Fig. \ref{fig5} the intrinsic activity is nevertheless noticeably masked, this time by mass transfer limitations in the boundary layer. At $u^{\rm inl} = 1$\,cm/s and $L = 1$\,cm the maximum observable TOFs are lower than the peak intrinsic ones, and a high-activity branch extends now to much more CO-poor feeds. As in the adiabatic limit there is a range of CO partial pressures for which we observe a bistability, and the results depend again quantitatively on the reactor setup. This is illustrated in Fig. \ref{fig5} by showing the data also for an increased ($L = 10$\,cm) and decreased ($L = 1$\,mm) inlet distance. For smaller inlet distances higher observable TOFs result. Highly comparable variations are obtained when changing the axial inlet velocity by one order of magnitude up or down, with smaller $u^{\rm inl}$ yielding larger observable TOFs (not shown). The varying reactor conditions also affect the extension
of the high-activity branch, which at maximum reaches down to $p^{\rm inl}_{\rm CO} = 0.6$\,atm, i.e with $p^{\rm inl}_{\rm O_2} = 0.3$\,atm to stoichiometric feed. In addition, there is now in principle also a dependence of the results on the employed sample thickness $d$, which enters through the boundary condition, eq. (23). However, due to the high thermal conductivity this dependence is in practice negligible. Compared to the results in Fig. \ref{fig5}, which were obtained using $d = 1$\,mm, changing the thickness by one order of magnitude up or down has virtually no effect on the observable catalytic activity.

\subsubsection{Mass transfer limitations}

The entire range of transport effects observed in the isothermal limit is this time due to mass transfer limitations in the boundary layer. At the high intrinsic TOFs around peak performance the mass conversion at the active surface is so large that these limitations lead to noticeable changes of the partial pressure profiles from inlet to surface. As schematically shown in Fig. \ref{fig3} there is essentially a build-up of a significant product concentration that is no longer sufficiently quickly removed. This goes hand in hand, cf. eq. (\ref{machcond}), with a decrease in reactant partial pressures, i.e. O$_2$ and CO are hindered in their access to the active surface. Due to the similar transport parameters and mass of both diatomic reactants, cf. Table \ref{tableI}, this limitation affects both species similarly. As long as the nominal inlet composition is different from stoichiometric feed, a corresponding roughly equal reduction of both reactant partial pressures close to the surface will then effectively change the $p^{\rm surf}_{\rm CO}/p^{\rm surf}_{\rm O_2}$ ratio. As also apparent from Fig. \ref{fig4} at $T^{\rm inl} = 600$\,K the range of peak intrinsic activity corresponds to quite CO-rich feeds. In this range the mass transfer limitations therefore lead to a noticeable increase of the $p^{\rm surf}_{\rm CO}/p^{\rm surf}_{\rm O_2}$ ratio compared to the nominal inlet composition as shown in Fig. \ref{fig5}. At a nominal inlet composition that would correspond to optimum intrinsic activity, $p^{\rm inl}_{\rm CO} \approx 3$\,atm in Fig. \ref{fig5}, the surface then sees a comparatively more CO-rich feed and the observable TOFs are lowered compared to the intrinsic ones. On the other hand, at a nominal inlet composition only slightly more CO-rich than stoichiometric feed, where the intrinsic activity would already have collapsed in Fig. \ref{fig5}, the significantly more CO-rich feed effectively seen by the surface corresponds in fact to conditions close to optimum intrinsic activity. The observable TOF is much increased and the high-activity branch of Fig. \ref{fig5} results. Exactly at stoichiometric feed this effective CO enrichment close to the surface ends, and for even more CO-poor mixtures possible mass transfer limitations would rather suppress the CO minority species. However, for such partial pressure ratios the intrinsic activity is low anyway and no mass transfer limitations arise. At the latest the high-activity branch therefore breaks down at stoichiometric feed. With this understanding the observed variations with the reactor setup are also easy to rationalize. A smaller boundary layer as resulting from increased axial inlet velocity or reduced inlet distance reduces the mass transfer limitations. The partial pressure ratio at the surface gets closer to the nominal one. In turn, the observable TOFs approach the intrinsic ones as illustrated in Fig. \ref{fig5} for the varying inlet distances.

\begin{figure}
\includegraphics[width=\linewidth]{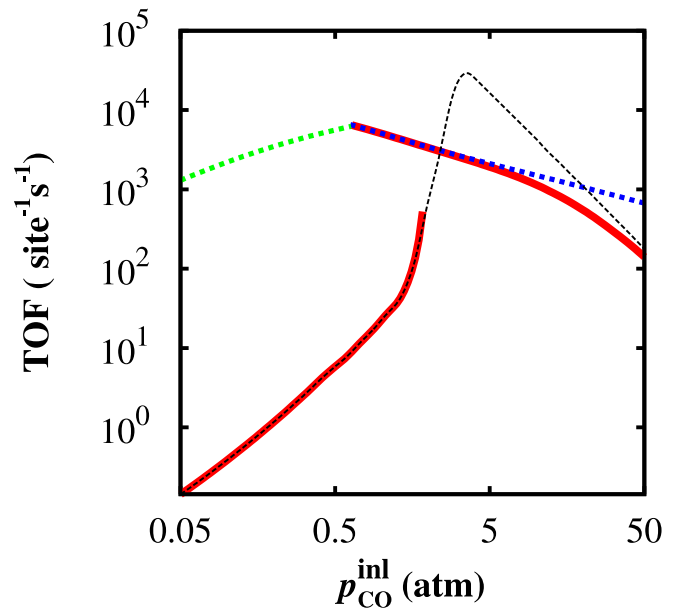}
\caption{(Color online) Same as Fig. \ref{fig5}, but including the upper TOF limit set entirely by mass transfer (dotted line), see text. The blue dotted line indicates the range limited by oxygen mass transfer, the green dotted line the range limited by CO mass transfer. Shown is data for $p^\text{inl}_\text{O$_2$} = 0.3$\,atm, $T^{\rm inl} = 600$\,K, $u^{\rm inl} = 1$\,cm/s, $L = 1$\,cm.}
\label{fig6}
\end{figure}
 
In the presence of such mass transfer limitations a natural question is to what degree they mask the intrinsic surface reactivity. Is the observable TOF profile the result of a complex mixture of the on-going surface chemistry and the gas-phase transport, or are the flow conditions in the reactor such that the measured activity conveys little information about the actual catalyst anymore. To qualify this it is useful to assess the upper TOF limit that results if mass flux is the only limitation. Such an estimate can be obtained by realizing that the steady-state mass conversion by the catalyst can never be higher than the one that completely depletes the minority species at the surface. Rather than using the catalyst specific boundary condition eq. (\ref{chemsource}) that depends on the actual intrinsic TOFs, we then simply employ for O-rich feeds
\begin{equation}
p^{\rm surf}_{\rm CO} = 0 \quad  ,
\end{equation}
and for CO-rich feeds
\begin{equation}
p^{\rm surf}_{\rm O_2} = 0 \quad . 
\end{equation}
For the respective majority species the boundary condition is still eq. (\ref{chemsource}), but the TOF entering this equation is now determined by the mass flux for the minority species, i.e. the conversion is completely dictated by the amount of impinging minority species. Figure \ref{fig6} shows the upper TOF limit that results from this estimate for the afore discussed gas-phase conditions of Fig. \ref{fig5}. Apparently, the observable TOFs come very close to this upper limit for most of the active region. This means that in this regime the measurable profile has very little to do with the actual RuO$_2$(110) catalyst, it rather reflects only the flow conditions in the employed reactor. Obviously, and similar to the adiabatic limit discussed before, if corresponding effects are not appropriately accounted for in {\em in situ} studies, wrong conclusions about the surface chemistry at technologically relevant gas-phase conditions will be derived.

\section{Summary and conclusions}

In summary we have presented an efficient approach to couple first-principles kinetic Monte Carlo and fluid dynamical simulations in the context of heterogeneous catalysis. This augments the accurate description of the surface chemistry achieved by 1p-kMC with a continuum account of the heat and mass transfer in a given reactor geometry, or vice versa it integrates the accurate 1p-kMC microkinetics into reactor-level modeling. In prevalent chemical engineering approaches the latter modeling instead incorporates phenomenological microkinetic treatments based on mean-field rate equations. In contrast to such a description, the presented 1p-kMC based multiscale modeling approach derives the average flux quantities required for the macroscopically described flow field properly from microscopic simulations that fully account for the site heterogeneity and distributions at the catalyst surface. As such it has the potential to carry the predictive power of the underlying electronic structure calculations for the elementary processes all the way to the reactor level.

On the way to such a first-principles chemical engineering we have applied the approach to the problem of {\em in situ} studies of model catalysts using the ambient pressure CO oxidation at RuO$_2$(110) as a representative example. As a suitable, though idealized reactor model to discuss the transport at the flat-faced surface we have chosen a stagnation flow geometry. The observed catalytic function depends sensitively on the employed reactor geometry (mimicked in this study by varying the inlet-surface distance) and the applied throughput rate (i.e. the streaming velocity at the inlet). For the thin, well heat conducting single crystal the degree of heat dissipation possible at the back of the sample (e.g. through radiative loss or contact to the sample holder) is a further crucial factor. Not aiming, nor being able to address specific experimental realizations this was considered through two opposite extremes: The isothermal limit mimicking a highly efficient heat coupling of the crystal to the system, and the adiabatic limit to represent a well insulated sample. In both limits transport effects were found to readily mask the intrinsic catalytic function at the high conversion rates reached at near-ambient gas-phase conditions. In the adiabatic limit this is due to a significant surface heating, in the isothermal limit due to mass transfer limitations that lead to the build-up of a significant concentration of products in the boundary layer above the active surface. With the single-crystal in real experimental setups neither perfectly heat-coupled nor isolated, these two effects discussed here separately will obviously be intricately intermingled and need to be disentangled by dedicated measurements and setups. Furthermore, we obtained in both limits a range of gas-phase conditions where the system exhibits two stationary operation modes, a low-activity branch corresponding to the intrinsic reactivity and a high-activity branch which arises from the coupling of the surface chemistry to the surrounding flow field. A corresponding bistability obtained here in the steady-state limit clearly suggests that the system could oscillate between the two modes, possibly even inhomogeneously in form of reaction fronts over the single-crystal surface. In case of heat transfer limitations, an intuitive propagation mechanism would hereby be via the formation of local hot spots, while in the mass transfer case it would be via gas-phase coupling, with the presented approach establishing the intriguing possibility to quantify these model conceptions with first-principles based simulations.

The main objective of {\em in situ} studies of model catalysts is a detailed, atomic-scale analysis of the catalytic function at technologically relevant gas-phase conditions, thereby bridging the pressure gap to the at present often much better characterized function in UHV. The range of transport effects discussed in the present study qualifies the additional complexity that needs to be accounted for in corresponding work to prevent wrong mechanistic conclusions about the surface chemistry at high pressure. That this complexity has potentially not yet been sufficiently appreciated may very well be the reason for the many existing controversies in the field. Also because of the limitation of the employed 1p-kMC RuO$_2$(110) model with respect to a reduction of the oxide catalyst we have refrained from comparing our simulations to already published experimental data. Nevertheless, we note that the gas-phase conditions and TOFs discussed here are of the order of magnitude presented in a manifold of {\em in situ} studies of CO oxidation at late transition metal catalysts. In this respect it is important to recognize that the CO oxidation reaction -- that has been a fruitfly reaction in UHV Surface Science due to its alleged "simplicity" and model character -- requires particular attention. The high turnovers that can be reached precisely because of this "simplicity" make this reaction much more prone to transport effects than other more complex, selective ones.

\begin{acknowledgments}

Funding within the MPG Innovation Initiative Multiscale Materials Modeling of Condensed Matter and the DFG Cluster of Excellence Unifying Concepts in Catalysis is gratefully acknowledged.

\end{acknowledgments}

\bibliography{prb_matera}

\end{document}